\documentclass[twoside,12pt]{article}

\usepackage{amsmath}
\usepackage{xspace}
\usepackage{boxedminipage}
\usepackage{alltt}
\usepackage{graphicx}
\usepackage{verbatim}
\usepackage{varioref}
\usepackage{pifont}
\usepackage{multicol}
\usepackage{theorem}

\newcommand{\GNU}        {{\small GNU}\xspace}

\newcommand{\UNIX}       {{\small UNIX}\xspace}

\newcommand{\WWW}        {{\small WWW}\xspace}
\newcommand{\GDB}        {{\small GDB}\xspace}

\newcommand{\PERL}       {{\small PERL}\xspace}
\newcommand{\STRIPE}     {{\small STRIPE}\xspace}

\newcommand{\PASS}{\text{\ding{52}}\xspace}
\newcommand{\FAIL}{\text{\ding{56}}\xspace}
\newcommand{\UNRESOLVED}{\lower0.30ex\hbox{\includegraphics*[height=1.7ex]{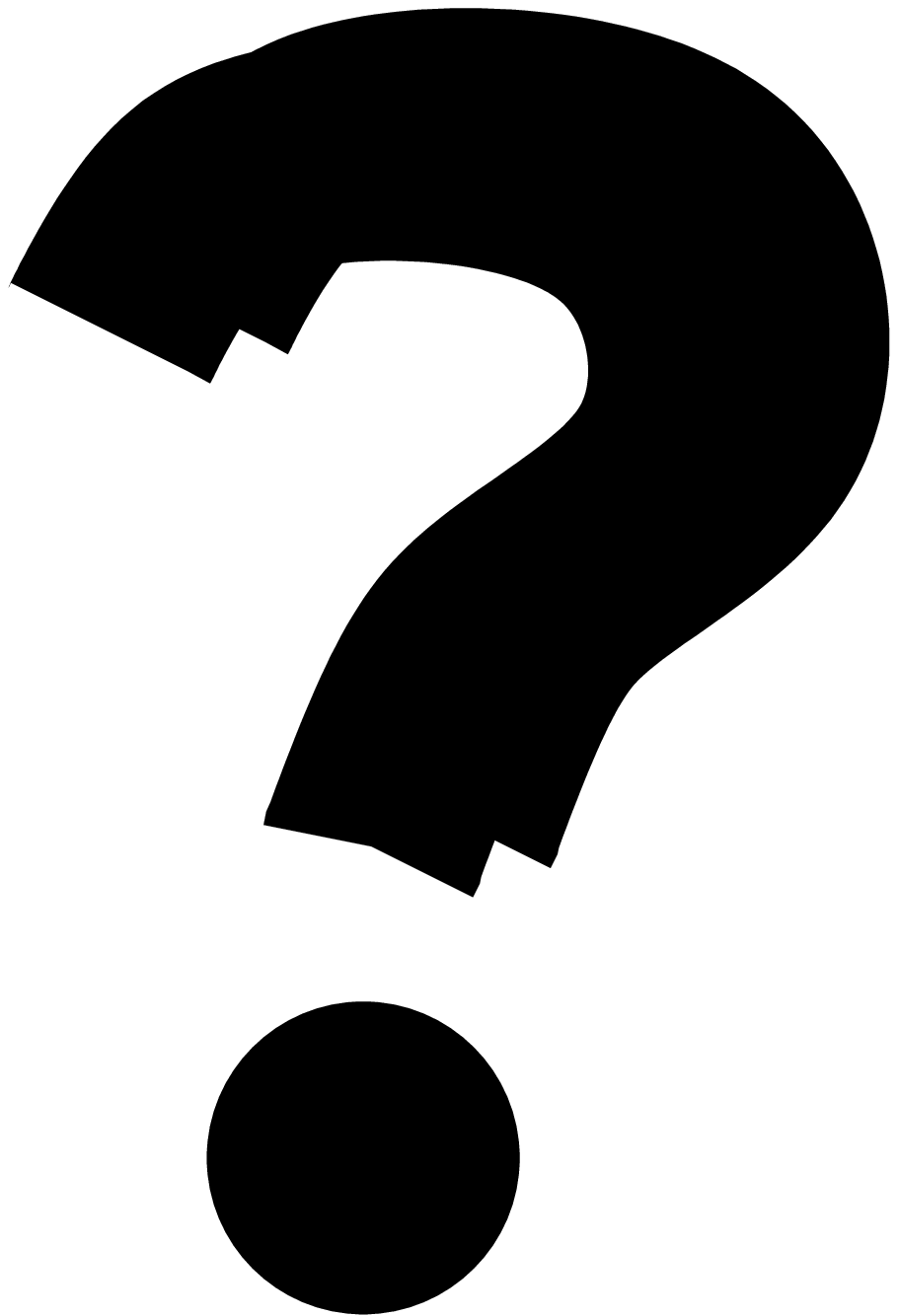}}}

\def\stripelog#1#2{\fontsize{3}{4pt}\selectfont #1 \quad #2}
\def\T{$\bullet$}
\def\F{\makebox[3.68pt]{.}}

\newcommand{\CC}{C}
\newcommand{\ddii}{\textit{dd}_2\xspace}
\newcommand{\edd}{\textit{dd}^+\xspace}
\newcommand{\ddmin}{\textit{ddmin}\xspace}

\newcommand{\test}{\textit{test}\xspace}

\newcommand{\dq}{\char34}

\newcommand{\mathid}[1]{\text{\rmfamily\textit{#1}}}

\def\|#1|{\mathid{#1}}

\newcommand{\codeid}[1]{\text{\upshape\texttt{#1}}}

\def\<#1>{\codeid{#1}}

\newtheorem{definition}{Definition}

\newtheorem{axiom}[definition]{Axiom}

\DeclareGraphicsExtensions{.ps,.eps,.pdf}

\begin{document}
\pagestyle{myheadings} 
\markboth{AADEBUG 2000}{Finding Failure Causes through Automated Testing} 
\title{Finding Failure Causes through~Automated~Testing\footnote{In 
    M. Ducass{\'e} (ed), 
    proceedings of the Fourth International Workshop on Automated
    Debugging (AADEBUG 2000), August 2000, Munich. COmputer Research Repository
    (http://www.acm.org/corr/), cs.SE/0012009; 
    whole proceedings: cs.SE/0010035.}}
\author{Holger Cleve \\ 
        Universit{\"a}t Passau\\
        Lehrstuhl Software-Systeme\\
        Innstra{\ss}e 33\\
        94032 Passau, Germany\\
        +49 851 509-3094\\
        cleve@fmi.uni-passau.de
        \and
        {Andreas Zeller}\\
        Universit{\"a}t Passau\\
        Lehrstuhl Software-Systeme\\
        Innstra{\ss}e 33\\
        94032 Passau, Germany\\
        +49 851 509-3095\\
        zeller@acm.org}
\date{}
\maketitle 

\begin{abstract}
A program fails.  Under which circumstances does this failure occur?
One single algorithm, the \emph{delta debugging} algorithm, suffices
to determine these \emph{failure-inducing circumstances.}  Delta
debugging tests a program systematically and automatically to isolate
failure-inducing circumstances such as the program input, changes to
the program code, or executed statements.
\medskip

\end{abstract}

\section{Debugging by Testing}
\label{sec:intro}

Debugging falls into three phases: reproducing a failure, finding the
root cause of the failure, and correcting the error such that the
failure no longer occurs.  While failure reproduction and correction
are important issues, it is the second phase, finding the root cause,
which is the most significant.  Early studies have shown that finding
the root cause accounts for 95\% of the whole debugging
effort~\cite{myers/79/testing}.

The common definition of a cause is \emph{some preceding event without
  which the effect would not have occurred}.  This implies that any
claim for causality can only be verified by experimentation.  To prove
that an event is the cause, the effect must no longer occur if the
event is missing.

In the context of debugging, the ``effect'' is a failure, and the
``event'' is some circumstance of the program execution.  To prove
that some circumstance has caused a failure, one must remove it in a
way that the failure no longer occurs.  This is typically done as the
very last debugging step: After correcting the error, one re-tests the
program, verifies that the failure is gone, and thus proves that the
original error was indeed the failure cause.

In order to prove causality, there can be no substitute for
testing---not even the most sophisticated analysis of the original
program run.  In other words: \emph{Analysis can show the absence of
  causality, but only testing can show its presence.}

In this paper, we give an overview on \emph{delta debugging}---an
automated debugging method that relies on systematic testing to prove
and isolate failure causes---circumstances such as the program input,
changes to the program code, or executed statements.  Basically, delta
debugging sets up subsets of the original circumstances, and tests
these configurations whether the failure still occurs.  Eventually,
delta debugging returns a subset of circumstances where every single
circumstance is relevant for producing the failure.

This paper unifies our previous work on delta
debugging~\cite{hildebrandt/zeller/2000/issta,zeller/99/esec} by
showing that a single algorithm suffices.  For the first time, the
isolation of failure-inducing statements is discussed.  After basic
definitions and a discussion of the algorithm, we show how to apply
delta debugging to identify failure-inducing changes, failure-inducing
program input, and failure-inducing statements.  We close with
discussions of related and future work.

\section{Configurations and Tests}
\label{sec:basics}

Let us begin with some basic definitions.  First of all, what are the
``circumstances'' of a failure?  Roughly speaking, a failure
circumstance is anything that might influence the existence of the
failure.  Without any further knowledge, this is anything that may
influence the program's execution: its environment, its input, its
code.  All these are circumstances of a failure.

We call a set of circumstances a \emph{scenario}.  Obviously, the root
cause of a problem is somewhere within this scenario.  To isolate the
root cause, we must separate the chaff from the wheat, or irrelevant
from relevant failure circumstances.

Let us now have a specific failing scenario to investigate.  Normally,
we also know a \emph{working scenario} under which the failure does
\emph{not} occur.  Let us assume we have some \emph{gradual
  transition} from the working scenario to the failing scenario---for
instance, by adding or altering circumstances.  The idea then is to
systematically test scenarios along this transition in order to
isolate failure-inducing circumstances and to use the test outcome to
generate new hypotheses.

Formally, we view a \emph{failure-inducing scenario} $\CC$ as the
result of applying a number of \emph{changes} $\Delta_1, \Delta_2,
\ldots, \Delta_n$ to some working scenario.  This way, we have a
gradual \emph{transition} from the working scenario (= no changes
applied) to $\CC$ (= all changes applied).

We can describe any scenario between the working scenario and $\CC$ as
a \emph{configuration of changes}:

\begin{definition}[Scenario]
  Let $\CC = \{\Delta_1, \Delta_2, \ldots, \Delta_n\}$ be a set of changes.  
  A change set $c \subseteq \CC$ is called a \emph{scenario.}
\end{definition}

\noindent
A scenario is constructed by applying changes to the working scenario:
\begin{definition}[Working scenario]
An empty scenario $c = \emptyset$ is called the \emph{working scenario.}
\end{definition}

We do not impose any constraints on how changes may be combined; in
particular, we do not assume that changes are ordered.  Thus, in the
worst case, there are $2^n$~possible scenarios for $n$~changes.

To determine whether a scenario fails, we assume a
\emph{testing function} with three standard outcomes:

\begin{definition}[Test]
The function $\test: 2^{\CC} \to \{\FAIL, \PASS, \UNRESOLVED\}$
determines for a scenario whether
some given failure occurs~(\FAIL) 
or not~(\PASS) 
or whether the test is unresolved~(\UNRESOLVED).
\end{definition}

\noindent
In practice, \test would construct the scenario by applying the given
changes to the working scenario, execute the scenario and return
the outcome.

\begin{figure*}[t]
\begin{boxedminipage}{\textwidth}\footnotesize
\subsection*{Minimizing Delta Debugging Algorithm}

The \emph{minimizing delta debugging algorithm} $\ddmin(c)$ is 
\begin{align*}
\ddmin(c) &= \ddmin_2(c, 2) \quad \text{where} \\
\ddmin_2(c, n) &= 
\begin{cases}
\ddmin_2(c_i, 2) & \text{if $\test(c_i) = \FAIL$ for some $i$} \\
  & \text{\qquad (``reduce to subset'')} \\
\ddmin_2\bigl(\bar{c_i}, \max(n - 1, 2)\bigr) & \text{else if $\test(\bar{c_i}) = \FAIL$ for some $i$} \\
  & \text{\qquad (``reduce to complement'')} \\
\ddmin_2\bigl(c, \min(|c|, 2n)\bigr) & \text{else if $n < |c|$} \\
  & \text{\qquad (``increase granularity'')} \\
c  & \text{otherwise (``done'').}
\end{cases}
\end{align*}
where $c_1, \ldots, c_n \subseteq c$ such that $\textstyle \bigcup c_i = c$, all
$c_i$ are pairwise disjoint, $\forall c_i \: (|c_i| \approx |c| / n)$, 
as~well~as $\bar{c_i} = c - c_i$.
\medskip

The recursion invariant (and thus precondition) for $\ddmin_2$ is 
$\test(c) = \FAIL \land n \leq |c|$.
\end{boxedminipage}
\caption{Minimizing delta debugging algorithm}
\label{fig:tcmin}
\end{figure*}

Let us now model our initial setting.  We have some \emph{working
  scenario} that works fine and some scenario that fails:

\begin{axiom}[Scenarios]  
  $\test(\emptyset) = \PASS$ (``working scenario'') and 
  $\test(\CC) = \FAIL$ (``failing scenario'') hold.
\end{axiom}

\noindent
Our goal is now to \emph{minimize} the failing scenario $\CC$---that
is, making it as similar as possible to the working scenario.  A
scenario~$c$ being ``minimal'' means that no subset of~$c$ causes the
test to fail.  Formally:

\begin{definition}[Minimal scenario]
\label{def:minimal}
A scenario $c \subseteq \CC$ is \emph{minimal} if
$
\forall c' \subset c \: \bigl(\test(c') \neq \FAIL\bigr)
$
holds.
\end{definition}

\noindent
This is what we want: to minimize a scenario such that \emph{all
  circumstances are relevant in producing the failure}---removing any
change causes the failure to disappear.

\section{Minimality of Scenarios}
\label{sec:algorithm}

How can one actually determine a minimal scenarios?  Here comes bad
news.  Let there be some scenario~$c$ consisting of $|c|$~changes to
the minimal scenario.  Relying on \test alone to determine minimality
requires testing all~$2^{|c|} - 1$ true subsets of $c$, which
obviously has exponential complexity.

What we can determine, however, is an \emph{approximation}---for
instance, a scenario where still every part on its own is significant
in producing the failure, but we do not check whether removing several
parts at once might make the scenario even smaller.  Formally, we
define this property as \mbox{\emph{1-minimality}}, where
$n$-minimality is defined as:

\begin{definition}[$n$-minimality]
\label{def:n-minimal}
A scenario $c \subseteq \CC$ is \emph{$n$-minimal} if
$
\forall c' \subset c \: \bigl( |c| - |c'| \leq n \Rightarrow \bigl(\test(c') \neq \FAIL\bigr)\bigr)
$
holds.
\end{definition}

\noindent
If $c$ is $|c|$-minimal, then $c$~is minimal in the sense of
Definition~\ref{def:minimal}.

Definition~\ref{def:n-minimal} gives a first idea of what we should be
aiming at.  However, given a scenario with, say, 100,000 changes, we
cannot simply undo each individual change in order to minimize it.
Thus, we need an effective algorithm to reduce our scenario efficiently.

An example of such a minimization algorithm is the \emph{minimizing
  delta debugging algorithm} $\ddmin$, shown in
Figure~\vref{fig:tcmin} and discussed
in~\cite{hildebrandt/zeller/2000/issta}.  $\ddmin$ is based on the
idea of \emph{divide and conquer.}  Initially, it partitions the set
of changes into two subsets $c_1$~and~$c_2$ and tests each of them: if
any test fails, the search is continued with this reduced subset.

If no test fails, $\ddmin$ \emph{increases the granularity} by
\emph{doubling} the number of subsets.  It then tests each subset and
each \emph{complement}; if the test fails, the search is continued
with this reduced subset.  The process is continued until each subset
has a size of~1 and no further reduction is possible.

\begin{figure}[t]
\includegraphics[width=\columnwidth]{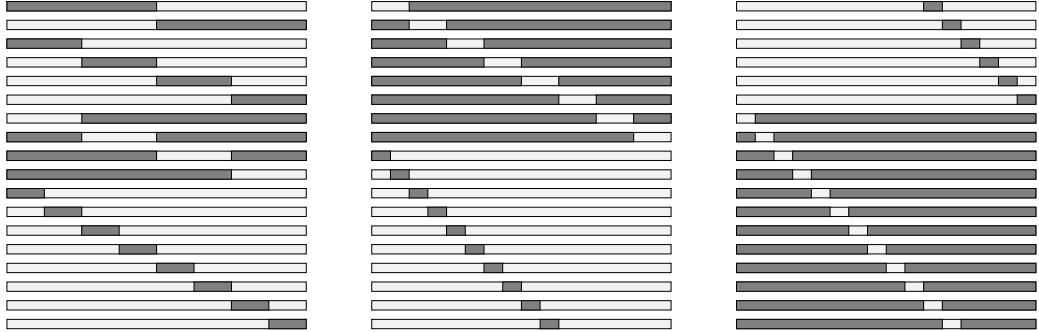}
\caption{Tests carried out by $\ddmin$}
\vspace{-0.25cm}
\label{fig:tcmin3}
\end{figure}

The whole process is illustrated in Figure~\vref{fig:tcmin3}.  Here,
we assume that every test outcome is unresolved.  We see how $\ddmin$
first partitions the whole set of changes (a rectangle) into two
subsets (gray areas), then into four, then into eight, sixteen, etc.,
testing each subset as well as its complement.

The $\ddmin$ algorithm guarantees that the returned set is 1-minimal;
that is, no single change that can be removed such that the test still
fails.  In the last stages of Figure~\ref{fig:tcmin3}, we see how this
guarantee is achieved: $\ddmin$ returns only when all complements have
passed the test, and this means that each remaining change has been
removed at least once.

This guarantee comes at a price.  In the worst case (every test fails
but the last complement), $\ddmin$ requires up to $n^2 + 3n$ tests for
$n$~changes~\cite{hildebrandt/zeller/2000/issta}.  However, in the
best case (every test fails), $\ddmin$ requires only $\log_2 n$ tests.

The performance of $\ddmin$ can be dramatically improved if we know
that once a set of changes passes the test, then every subset passes
the test as well---the so-called \emph{monotony.}  If we know that a
set of changes is monotone, we need not test sets whose supersets have
already passed the test: $\test$ can simply return $\PASS$ without
actually carrying out the test.

This optimization makes $\ddmin$ linear at worst---but only if there
are no unresolved test outcomes.  The number of tests required by
$\ddmin$ largely depends on our ability to group the scenario
transitions such that we avoid unresolved test outcomes and increase
our chances to get failing tests.

\section{Finding Failure-Inducing Changes}
\label{sec:dd}

In our first case study~\cite{zeller/99/esec}, we applied delta
debugging to a common, yet simple regression case.  The situation is
as follows: There is some old version of the program (``yesterday''),
which works, and a new version (``today'') which does not.  The goal
is to identify the changes to the program code which induce the
failure.

This situation fits nicely into the scenario model of
Section~\ref{sec:basics}.  The ``yesterday'' version is the working
scenario; the changes are textual changes to the program code; the
``today'' version is the failing scenario where all changes are
applied.  Using the scenario minimization algorithm, we should be able
to minimize the set of applied changes and thus identify a small set
of failure-inducing changes.\footnote{In this first case study, we
  actually used a variant of $\ddmin$, called
  $\edd$~\cite{zeller/99/esec}.  For the examples in this Section,
  $\edd$ has exactly the same performance as $\ddmin$ under monotony;
  i.e.\ $\test(c)$ returns $\PASS$ without actually testing~$c$ if a
  superset of~$c$ has already passed the test.  $\ddmin$ does not
  guarantee 1-minimality, which is why we recommend $\ddmin$ (possibly
  with the monotony optimization), as a general replacement for
  $\edd$.}

Our results with delta debugging were quite promising.  First, if the
dependency between changes was known (from a version history, for
instance), then delta debugging became quite trivial: If every change
$\Delta_i$ depended on earlier changes $\Delta_1, \dots, \Delta_{i-1}$, then every
scenario which would not fulfill these constraints could be rejected
straight away.  Under these conditions, only configurations with a
full set of changes $\Delta_1, \dots, \Delta_{i-1}$ actually had to be
tested---$\ddmin$ degraded into a simple \emph{binary search} with
logarithmic complexity.

But even when the dependency between changes was not known, or when a
single logical change was still too large, $\ddmin$ performed
satisfactorily.  In one case study, we had a single 178,000-line
change to the \GNU debugger (\GDB); this change was broken down into
8721~textual changes in the \GDB source, with any two textual changes
separated by a context of at least two unchanged lines.  The problem
was, of course, that applying any subset of these 8721~changes did not
have many chances to result in something consistent.

\begin{figure}
\includegraphics[width=\columnwidth]{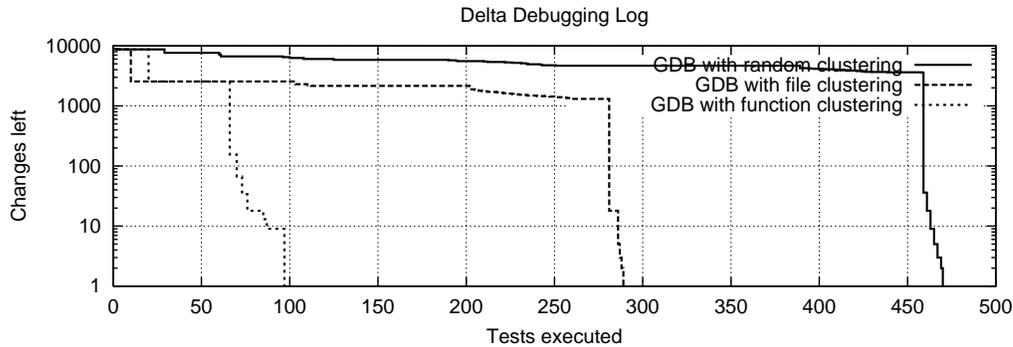}
\caption{Simplifying failure-inducing \GDB changes}
\vspace{-0.3cm}
\label{fig:gdb}
\end{figure}

To summarize: Nearly all tests were unresolved.  $\ddmin$ had a chance
to succeed only after the subsets and their complements had become
sufficiently close to $\emptyset$ and $\CC$, respectively.  As shown
in Figure~\vref{fig:gdb}, $\ddmin$ required 470~tests to isolate the
single failure-inducing change.  Each test involved change
application, smart reconstruction of \GDB, and running the test, which
took an average time of 190~seconds.\footnote{All times were measured
  on a Linux PC with a 500 MHz Pentium~III processor.}

After grouping changes by location criteria (i.e. common files and
directories), $\ddmin$ required 289~tests.  After applying syntactic
criteria (grouping of changes according to functions), $\ddmin$
required only~97 tests, or about 4 hours.

\section{Simplifying Test Cases}
\label{sec:tcmin}

In a second case study~\cite{hildebrandt/zeller/2000/issta}, we used
delta debugging to simplify \emph{failing test cases}---by minimizing
the differences between the failing input and a trivial (empty) input.

Having a minimal test case is an important issue for debugging.  Not
only does it show the relevant failure circumstances; simplifying test
cases also helps to identify and eliminate duplicate bug reports.  It
turned out that $\ddmin$ works just as well to minimize failure-inducing
input than to minimize failure-inducing changes.

Here comes another case study.  \emph{Mozilla,} Netscape's new web
browser project~\cite{mozilla.org}, is still in beta status and thus
shows numerous bugs.  We hunted a bug in which printing a certain \WWW
page caused the browser to fail.  This scenario gave us two things to
look for:

\begin{itemize}
\item Which parts of the \WWW page are relevant in reproducing the failure?
\item Which user actions are relevant in reproducing the failure?
\end{itemize}

\noindent
Using delta debugging, we could successfully simplify both the
sequence of user actions as well as the \WWW page.  It turned out that
a number of user actions (such as changing the printer settings) were
not relevant in reproducing the failure.  As shown in
Figure~\vref{fig:mozilla}, our prototype minimized 95~user actions
into~3 after 39~test runs (or 25 minutes): In order to crash Mozilla,
it suffices to press \emph{Alt+P} and press and release the mouse
button on \emph{Print}.\footnote{Note that releasing the \emph{P} key
  is not required.}  Likewise, in 57~test runs, our prototype reduced
the \WWW page from 896~lines to a single line:
\begin{alltt}
<SELECT NAME="priority" MULTIPLE SIZE=7>
\end{alltt}

\begin{figure}[t]
\vspace{-1.25cm}
\includegraphics*[width=\columnwidth]{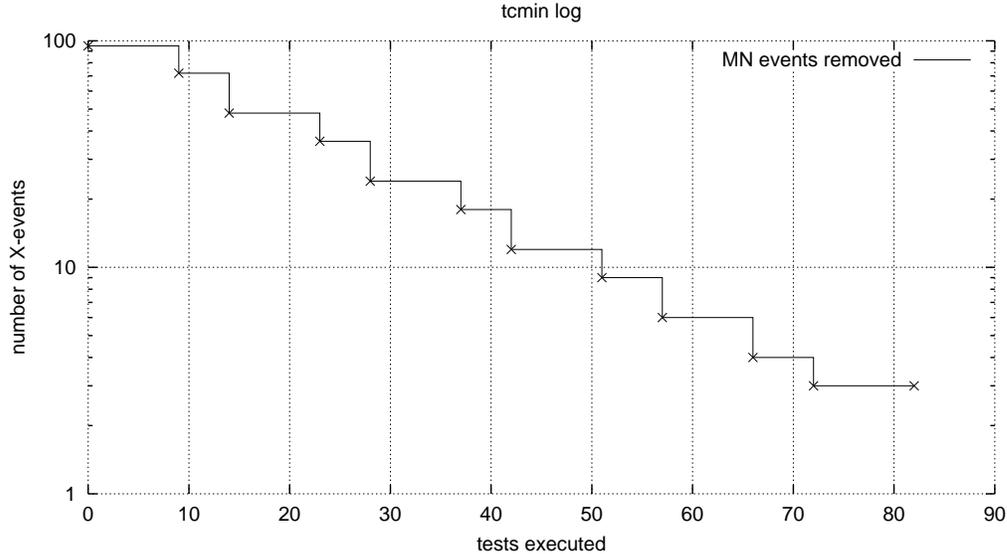}
\caption{Simplifying failure-inducing Mozilla input}
\vspace{-0.3cm}
\label{fig:mozilla}
\end{figure}

Automatic test case minimization was also applied to the \GNU C
compiler as well as various \UNIX utilities, all with promising
results.  All one needs is an input, a sequence of user actions, an
observable failure---and a little time to let the computer do the
minimization.

\section{Reducing Execution Traces}
\label{sec:tracered}
\enlargethispage{\baselineskip}

As a further example, we show how delta debugging can be helpful in
identifying \emph{failure-inducing events} during the execution of the
program.  We assume an \emph{execution trace} as a sequence of
\emph{program states} during execution, starting with a working
(empty) state and ending in a failing state.  During a program run,
statements are executed that access and alter the program state.  The
goal is now to identify those events (i.e.\ statement executions) that
were relevant in producing the failure and to eliminate those that
were not.

Again, this nicely fits into our scenario model of
Section~\ref{sec:basics}.  The ``working'' scenario is no execution at
all.  The ``failing'' scenario is the result of the state changes $\CC
= \{\Delta_1, \dots, \Delta_n\}$ induced by the executed statements.
Applying delta debugging to minimize~$\CC$ means to isolate exactly
those state changes that were relevant in producing the final
state---very much like well-known dynamic slicing
techniques~\cite{agrawal/horgan/90/pldi, korel/laski/90/jss,
  softools:tip}, but relying on \emph{partial execution} rather than
analysis of control and data flow and thus showing real causality
instead of potential causality.

As an example, consider the following \PERL program.  It reads in two
numbers $a$~and~$b$ and computes the sum $\|sum| = \sum_{i=a}^bi$ as
well as the product $\|mul| = \prod_{i=a}^bi$:

\def\N#1{{\makebox[7pt][r]{#1}\quad}}
\begin{quote}\small
\begin{alltt}\rmfamily
\N{ 1} \$\|sum| = 0;
\N{ 2} \$\|mul| = 1;
\N{ 3} print \texttt{\dq{}a? \dq{}}; \(\$a = \langle\rangle\);
\N{ 4} print \texttt{\dq{}b? \dq{}}; \(\$b = \langle\rangle\);
\N{ 5} \textbf{while} \((\$a \leq \$b)\) \{
\N{ 6}     \(\$\|sum| = \$\|sum| + \$a\);
\N{ 7}     \(\$\|mul| = \$\|mul| * \$a\);
\N{ 8}     \(\$a = \$a + 1\);
\N{ 9} \}
\N{10} print \texttt{\dq{}sum = \dq{}}, \$\|sum|, \texttt{\dq{}\(\backslash\)n\dq{}};
\N{11} print \texttt{\dq{}mul = \dq{}}, \$\|mul|, \texttt{\dq{}\(\backslash\)n\dq{}};
\end{alltt}
\end{quote}

\noindent
Here is an example run of \<sample.pl>:

\begin{quote}\small
\begin{alltt}
\$ \textbf{perl ./sample.pl}
a? \textbf{0}
b? \textbf{5}
sum = 15
mul = 0
\$ \_
\end{alltt}
\end{quote}

\noindent
For this run, we wanted to determine the events which have influenced
the final program output.  We set up a prototype called
\STRIPE{}\footnote{STRIPE = ``Systematic Trace Reduction by Iterative
  Partial Execution''} which applies the $\ddmin$ algorithm on
execution traces.

In a first step, \STRIPE determines the execution trace---that is, the
values of the program counter during execution.  Then, \STRIPE runs
the $\ddmin$ algorithm on the execution trace.  To omit a
statement~$S$ from execution, \STRIPE uses a conventional interactive
debugger to insert a breakpoint at~$S$.  The breakpoint causes the
program to interrupt whenever~$S$ is reached; an associated breakpoint
command causes the debugger to resume execution behind~$S$.

\begin{figure*}
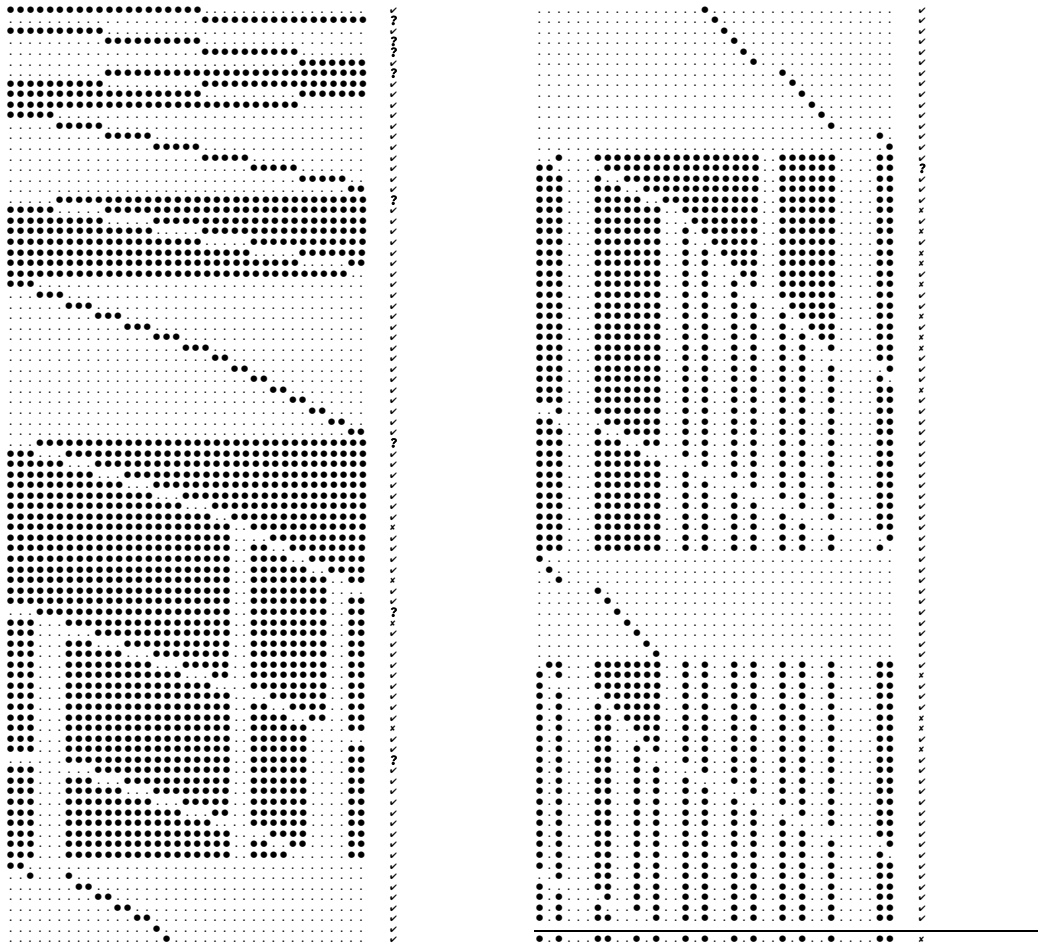

\begin{multicols}{2}
\stripelog{\T\T\T\T\T\T\T\T\T\T\T\T\T\T\T\T\T\T\T\T\F\F\F\F\F\F\F\F\F\F\F\F\F\F\F\F\F}{\PASS}

\stripelog{\F\F\F\F\F\F\F\F\F\F\F\F\F\F\F\F\F\F\F\F\T\T\T\T\T\T\T\T\T\T\T\T\T\T\T\T\T}{\UNRESOLVED}

\stripelog{\T\T\T\T\T\T\T\T\T\T\F\F\F\F\F\F\F\F\F\F\F\F\F\F\F\F\F\F\F\F\F\F\F\F\F\F\F}{\PASS}

\stripelog{\F\F\F\F\F\F\F\F\F\F\T\T\T\T\T\T\T\T\T\T\F\F\F\F\F\F\F\F\F\F\F\F\F\F\F\F\F}{\UNRESOLVED}

\stripelog{\F\F\F\F\F\F\F\F\F\F\F\F\F\F\F\F\F\F\F\F\T\T\T\T\T\T\T\T\T\T\F\F\F\F\F\F\F}{\UNRESOLVED}

\stripelog{\F\F\F\F\F\F\F\F\F\F\F\F\F\F\F\F\F\F\F\F\F\F\F\F\F\F\F\F\F\F\T\T\T\T\T\T\T}{\PASS}

\stripelog{\F\F\F\F\F\F\F\F\F\F\T\T\T\T\T\T\T\T\T\T\T\T\T\T\T\T\T\T\T\T\T\T\T\T\T\T\T}{\UNRESOLVED}

\stripelog{\T\T\T\T\T\T\T\T\T\T\F\F\F\F\F\F\F\F\F\F\T\T\T\T\T\T\T\T\T\T\T\T\T\T\T\T\T}{\PASS}

\stripelog{\T\T\T\T\T\T\T\T\T\T\T\T\T\T\T\T\T\T\T\T\F\F\F\F\F\F\F\F\F\F\T\T\T\T\T\T\T}{\PASS}

\stripelog{\T\T\T\T\T\T\T\T\T\T\T\T\T\T\T\T\T\T\T\T\T\T\T\T\T\T\T\T\T\T\F\F\F\F\F\F\F}{\PASS}

\stripelog{\T\T\T\T\T\F\F\F\F\F\F\F\F\F\F\F\F\F\F\F\F\F\F\F\F\F\F\F\F\F\F\F\F\F\F\F\F}{\PASS}

\stripelog{\F\F\F\F\F\T\T\T\T\T\F\F\F\F\F\F\F\F\F\F\F\F\F\F\F\F\F\F\F\F\F\F\F\F\F\F\F}{\PASS}

\stripelog{\F\F\F\F\F\F\F\F\F\F\T\T\T\T\T\F\F\F\F\F\F\F\F\F\F\F\F\F\F\F\F\F\F\F\F\F\F}{\PASS}

\stripelog{\F\F\F\F\F\F\F\F\F\F\F\F\F\F\F\T\T\T\T\T\F\F\F\F\F\F\F\F\F\F\F\F\F\F\F\F\F}{\PASS}

\stripelog{\F\F\F\F\F\F\F\F\F\F\F\F\F\F\F\F\F\F\F\F\T\T\T\T\T\F\F\F\F\F\F\F\F\F\F\F\F}{\PASS}

\stripelog{\F\F\F\F\F\F\F\F\F\F\F\F\F\F\F\F\F\F\F\F\F\F\F\F\F\T\T\T\T\T\F\F\F\F\F\F\F}{\PASS}

\stripelog{\F\F\F\F\F\F\F\F\F\F\F\F\F\F\F\F\F\F\F\F\F\F\F\F\F\F\F\F\F\F\T\T\T\T\T\F\F}{\PASS}

\stripelog{\F\F\F\F\F\F\F\F\F\F\F\F\F\F\F\F\F\F\F\F\F\F\F\F\F\F\F\F\F\F\F\F\F\F\F\T\T}{\PASS}

\stripelog{\F\F\F\F\F\T\T\T\T\T\T\T\T\T\T\T\T\T\T\T\T\T\T\T\T\T\T\T\T\T\T\T\T\T\T\T\T}{\UNRESOLVED}

\stripelog{\T\T\T\T\T\F\F\F\F\F\T\T\T\T\T\T\T\T\T\T\T\T\T\T\T\T\T\T\T\T\T\T\T\T\T\T\T}{\PASS}

\stripelog{\T\T\T\T\T\T\T\T\T\T\F\F\F\F\F\T\T\T\T\T\T\T\T\T\T\T\T\T\T\T\T\T\T\T\T\T\T}{\PASS}

\stripelog{\T\T\T\T\T\T\T\T\T\T\T\T\T\T\T\F\F\F\F\F\T\T\T\T\T\T\T\T\T\T\T\T\T\T\T\T\T}{\PASS}

\stripelog{\T\T\T\T\T\T\T\T\T\T\T\T\T\T\T\T\T\T\T\T\F\F\F\F\F\T\T\T\T\T\T\T\T\T\T\T\T}{\PASS}

\stripelog{\T\T\T\T\T\T\T\T\T\T\T\T\T\T\T\T\T\T\T\T\T\T\T\T\T\F\F\F\F\F\T\T\T\T\T\T\T}{\PASS}

\stripelog{\T\T\T\T\T\T\T\T\T\T\T\T\T\T\T\T\T\T\T\T\T\T\T\T\T\T\T\T\T\T\F\F\F\F\F\T\T}{\PASS}

\stripelog{\T\T\T\T\T\T\T\T\T\T\T\T\T\T\T\T\T\T\T\T\T\T\T\T\T\T\T\T\T\T\T\T\T\T\T\F\F}{\PASS}

\stripelog{\T\T\T\F\F\F\F\F\F\F\F\F\F\F\F\F\F\F\F\F\F\F\F\F\F\F\F\F\F\F\F\F\F\F\F\F\F}{\PASS}

\stripelog{\F\F\F\T\T\T\F\F\F\F\F\F\F\F\F\F\F\F\F\F\F\F\F\F\F\F\F\F\F\F\F\F\F\F\F\F\F}{\PASS}

\stripelog{\F\F\F\F\F\F\T\T\T\F\F\F\F\F\F\F\F\F\F\F\F\F\F\F\F\F\F\F\F\F\F\F\F\F\F\F\F}{\PASS}

\stripelog{\F\F\F\F\F\F\F\F\F\T\T\T\F\F\F\F\F\F\F\F\F\F\F\F\F\F\F\F\F\F\F\F\F\F\F\F\F}{\PASS}

\stripelog{\F\F\F\F\F\F\F\F\F\F\F\F\T\T\T\F\F\F\F\F\F\F\F\F\F\F\F\F\F\F\F\F\F\F\F\F\F}{\PASS}

\stripelog{\F\F\F\F\F\F\F\F\F\F\F\F\F\F\F\T\T\T\F\F\F\F\F\F\F\F\F\F\F\F\F\F\F\F\F\F\F}{\PASS}

\stripelog{\F\F\F\F\F\F\F\F\F\F\F\F\F\F\F\F\F\F\T\T\T\F\F\F\F\F\F\F\F\F\F\F\F\F\F\F\F}{\PASS}

\stripelog{\F\F\F\F\F\F\F\F\F\F\F\F\F\F\F\F\F\F\F\F\F\T\T\F\F\F\F\F\F\F\F\F\F\F\F\F\F}{\PASS}

\stripelog{\F\F\F\F\F\F\F\F\F\F\F\F\F\F\F\F\F\F\F\F\F\F\F\T\T\F\F\F\F\F\F\F\F\F\F\F\F}{\PASS}

\stripelog{\F\F\F\F\F\F\F\F\F\F\F\F\F\F\F\F\F\F\F\F\F\F\F\F\F\T\T\F\F\F\F\F\F\F\F\F\F}{\PASS}

\stripelog{\F\F\F\F\F\F\F\F\F\F\F\F\F\F\F\F\F\F\F\F\F\F\F\F\F\F\F\T\T\F\F\F\F\F\F\F\F}{\PASS}

\stripelog{\F\F\F\F\F\F\F\F\F\F\F\F\F\F\F\F\F\F\F\F\F\F\F\F\F\F\F\F\F\T\T\F\F\F\F\F\F}{\PASS}

\stripelog{\F\F\F\F\F\F\F\F\F\F\F\F\F\F\F\F\F\F\F\F\F\F\F\F\F\F\F\F\F\F\F\T\T\F\F\F\F}{\PASS}

\stripelog{\F\F\F\F\F\F\F\F\F\F\F\F\F\F\F\F\F\F\F\F\F\F\F\F\F\F\F\F\F\F\F\F\F\T\T\F\F}{\PASS}

\stripelog{\F\F\F\F\F\F\F\F\F\F\F\F\F\F\F\F\F\F\F\F\F\F\F\F\F\F\F\F\F\F\F\F\F\F\F\T\T}{\PASS}

\stripelog{\F\F\F\T\T\T\T\T\T\T\T\T\T\T\T\T\T\T\T\T\T\T\T\T\T\T\T\T\T\T\T\T\T\T\T\T\T}{\UNRESOLVED}

\stripelog{\T\T\T\F\F\F\T\T\T\T\T\T\T\T\T\T\T\T\T\T\T\T\T\T\T\T\T\T\T\T\T\T\T\T\T\T\T}{\PASS}

\stripelog{\T\T\T\T\T\T\F\F\F\T\T\T\T\T\T\T\T\T\T\T\T\T\T\T\T\T\T\T\T\T\T\T\T\T\T\T\T}{\PASS}

\stripelog{\T\T\T\T\T\T\T\T\T\F\F\F\T\T\T\T\T\T\T\T\T\T\T\T\T\T\T\T\T\T\T\T\T\T\T\T\T}{\PASS}

\stripelog{\T\T\T\T\T\T\T\T\T\T\T\T\F\F\F\T\T\T\T\T\T\T\T\T\T\T\T\T\T\T\T\T\T\T\T\T\T}{\PASS}

\stripelog{\T\T\T\T\T\T\T\T\T\T\T\T\T\T\T\F\F\F\T\T\T\T\T\T\T\T\T\T\T\T\T\T\T\T\T\T\T}{\PASS}

\stripelog{\T\T\T\T\T\T\T\T\T\T\T\T\T\T\T\T\T\T\F\F\F\T\T\T\T\T\T\T\T\T\T\T\T\T\T\T\T}{\PASS}

\stripelog{\T\T\T\T\T\T\T\T\T\T\T\T\T\T\T\T\T\T\T\T\T\F\F\T\T\T\T\T\T\T\T\T\T\T\T\T\T}{\PASS}

\stripelog{\T\T\T\T\T\T\T\T\T\T\T\T\T\T\T\T\T\T\T\T\T\T\T\F\F\T\T\T\T\T\T\T\T\T\T\T\T}{\FAIL}

\stripelog{\T\T\T\T\T\T\T\T\T\T\T\T\T\T\T\T\T\T\T\T\T\T\T\F\F\F\F\T\T\T\T\T\T\T\T\T\T}{\PASS}

\stripelog{\T\T\T\T\T\T\T\T\T\T\T\T\T\T\T\T\T\T\T\T\T\T\T\F\F\T\T\F\F\T\T\T\T\T\T\T\T}{\PASS}

\stripelog{\T\T\T\T\T\T\T\T\T\T\T\T\T\T\T\T\T\T\T\T\T\T\T\F\F\T\T\T\T\F\F\T\T\T\T\T\T}{\PASS}

\stripelog{\T\T\T\T\T\T\T\T\T\T\T\T\T\T\T\T\T\T\T\T\T\T\T\F\F\T\T\T\T\T\T\F\F\T\T\T\T}{\PASS}

\stripelog{\T\T\T\T\T\T\T\T\T\T\T\T\T\T\T\T\T\T\T\T\T\T\T\F\F\T\T\T\T\T\T\T\T\F\F\T\T}{\FAIL}

\stripelog{\T\T\T\T\T\T\T\T\T\T\T\T\T\T\T\T\T\T\T\T\T\T\T\F\F\T\T\T\T\T\T\T\T\F\F\F\F}{\PASS}

\stripelog{\T\T\T\T\T\T\T\T\T\T\T\T\T\T\T\T\T\T\T\T\T\T\T\F\F\T\T\T\T\T\T\T\T\F\F\T\T}{\PASS}

\stripelog{\F\F\F\T\T\T\T\T\T\T\T\T\T\T\T\T\T\T\T\T\T\T\T\F\F\T\T\T\T\T\T\T\T\F\F\T\T}{\UNRESOLVED}

\stripelog{\T\T\T\F\F\F\T\T\T\T\T\T\T\T\T\T\T\T\T\T\T\T\T\F\F\T\T\T\T\T\T\T\T\F\F\T\T}{\FAIL}

\stripelog{\T\T\T\F\F\F\F\F\F\T\T\T\T\T\T\T\T\T\T\T\T\T\T\F\F\T\T\T\T\T\T\T\T\F\F\T\T}{\PASS}

\stripelog{\T\T\T\F\F\F\T\T\T\F\F\F\T\T\T\T\T\T\T\T\T\T\T\F\F\T\T\T\T\T\T\T\T\F\F\T\T}{\PASS}

\stripelog{\T\T\T\F\F\F\T\T\T\T\T\T\F\F\F\T\T\T\T\T\T\T\T\F\F\T\T\T\T\T\T\T\T\F\F\T\T}{\PASS}

\stripelog{\T\T\T\F\F\F\T\T\T\T\T\T\T\T\T\F\F\F\T\T\T\T\T\F\F\T\T\T\T\T\T\T\T\F\F\T\T}{\PASS}

\stripelog{\T\T\T\F\F\F\T\T\T\T\T\T\T\T\T\T\T\T\F\F\F\T\T\F\F\T\T\T\T\T\T\T\T\F\F\T\T}{\PASS}

\stripelog{\T\T\T\F\F\F\T\T\T\T\T\T\T\T\T\T\T\T\T\T\T\F\F\F\F\T\T\T\T\T\T\T\T\F\F\T\T}{\PASS}

\stripelog{\T\T\T\F\F\F\T\T\T\T\T\T\T\T\T\T\T\T\T\T\T\T\T\F\F\F\F\T\T\T\T\T\T\F\F\T\T}{\PASS}

\stripelog{\T\T\T\F\F\F\T\T\T\T\T\T\T\T\T\T\T\T\T\T\T\T\T\F\F\T\T\F\F\T\T\T\T\F\F\T\T}{\PASS}

\stripelog{\T\T\T\F\F\F\T\T\T\T\T\T\T\T\T\T\T\T\T\T\T\T\T\F\F\T\T\T\T\F\F\T\T\F\F\T\T}{\PASS}

\stripelog{\T\T\T\F\F\F\T\T\T\T\T\T\T\T\T\T\T\T\T\T\T\T\T\F\F\T\T\T\T\T\T\F\F\F\F\T\T}{\FAIL}

\stripelog{\T\T\T\F\F\F\T\T\T\T\T\T\T\T\T\T\T\T\T\T\T\T\T\F\F\T\T\T\T\T\T\F\F\F\F\F\F}{\PASS}

\stripelog{\T\T\T\F\F\F\T\T\T\T\T\T\T\T\T\T\T\T\T\T\T\T\T\F\F\T\T\T\T\T\T\F\F\F\F\T\T}{\PASS}

\stripelog{\F\F\F\F\F\F\T\T\T\T\T\T\T\T\T\T\T\T\T\T\T\T\T\F\F\T\T\T\T\T\T\F\F\F\F\T\T}{\UNRESOLVED}

\stripelog{\T\T\T\F\F\F\F\F\F\T\T\T\T\T\T\T\T\T\T\T\T\T\T\F\F\T\T\T\T\T\T\F\F\F\F\T\T}{\PASS}

\stripelog{\T\T\T\F\F\F\T\T\T\F\F\F\T\T\T\T\T\T\T\T\T\T\T\F\F\T\T\T\T\T\T\F\F\F\F\T\T}{\PASS}

\stripelog{\T\T\T\F\F\F\T\T\T\T\T\T\F\F\F\T\T\T\T\T\T\T\T\F\F\T\T\T\T\T\T\F\F\F\F\T\T}{\PASS}

\stripelog{\T\T\T\F\F\F\T\T\T\T\T\T\T\T\T\F\F\F\T\T\T\T\T\F\F\T\T\T\T\T\T\F\F\F\F\T\T}{\PASS}

\stripelog{\T\T\T\F\F\F\T\T\T\T\T\T\T\T\T\T\T\T\F\F\F\T\T\F\F\T\T\T\T\T\T\F\F\F\F\T\T}{\PASS}

\stripelog{\T\T\T\F\F\F\T\T\T\T\T\T\T\T\T\T\T\T\T\T\T\F\F\F\F\T\T\T\T\T\T\F\F\F\F\T\T}{\PASS}

\stripelog{\T\T\T\F\F\F\T\T\T\T\T\T\T\T\T\T\T\T\T\T\T\T\T\F\F\F\F\T\T\T\T\F\F\F\F\T\T}{\PASS}

\stripelog{\T\T\T\F\F\F\T\T\T\T\T\T\T\T\T\T\T\T\T\T\T\T\T\F\F\T\T\F\F\T\T\F\F\F\F\T\T}{\PASS}

\stripelog{\T\T\T\F\F\F\T\T\T\T\T\T\T\T\T\T\T\T\T\T\T\T\T\F\F\T\T\T\T\F\F\F\F\F\F\T\T}{\PASS}

\stripelog{\T\T\F\F\F\F\F\F\F\F\F\F\F\F\F\F\F\F\F\F\F\F\F\F\F\F\F\F\F\F\F\F\F\F\F\F\F}{\PASS}

\stripelog{\F\F\T\F\F\F\T\F\F\F\F\F\F\F\F\F\F\F\F\F\F\F\F\F\F\F\F\F\F\F\F\F\F\F\F\F\F}{\PASS}

\stripelog{\F\F\F\F\F\F\F\T\T\F\F\F\F\F\F\F\F\F\F\F\F\F\F\F\F\F\F\F\F\F\F\F\F\F\F\F\F}{\PASS}

\stripelog{\F\F\F\F\F\F\F\F\F\T\T\F\F\F\F\F\F\F\F\F\F\F\F\F\F\F\F\F\F\F\F\F\F\F\F\F\F}{\PASS}

\stripelog{\F\F\F\F\F\F\F\F\F\F\F\T\T\F\F\F\F\F\F\F\F\F\F\F\F\F\F\F\F\F\F\F\F\F\F\F\F}{\PASS}

\stripelog{\F\F\F\F\F\F\F\F\F\F\F\F\F\T\T\F\F\F\F\F\F\F\F\F\F\F\F\F\F\F\F\F\F\F\F\F\F}{\PASS}

\stripelog{\F\F\F\F\F\F\F\F\F\F\F\F\F\F\F\T\F\F\F\F\F\F\F\F\F\F\F\F\F\F\F\F\F\F\F\F\F}{\PASS}

\stripelog{\F\F\F\F\F\F\F\F\F\F\F\F\F\F\F\F\T\F\F\F\F\F\F\F\F\F\F\F\F\F\F\F\F\F\F\F\F}{\PASS}

\stripelog{\F\F\F\F\F\F\F\F\F\F\F\F\F\F\F\F\F\T\F\F\F\F\F\F\F\F\F\F\F\F\F\F\F\F\F\F\F}{\PASS}

\stripelog{\F\F\F\F\F\F\F\F\F\F\F\F\F\F\F\F\F\F\T\F\F\F\F\F\F\F\F\F\F\F\F\F\F\F\F\F\F}{\PASS}

\stripelog{\F\F\F\F\F\F\F\F\F\F\F\F\F\F\F\F\F\F\F\T\F\F\F\F\F\F\F\F\F\F\F\F\F\F\F\F\F}{\PASS}

\stripelog{\F\F\F\F\F\F\F\F\F\F\F\F\F\F\F\F\F\F\F\F\T\F\F\F\F\F\F\F\F\F\F\F\F\F\F\F\F}{\PASS}

\stripelog{\F\F\F\F\F\F\F\F\F\F\F\F\F\F\F\F\F\F\F\F\F\T\F\F\F\F\F\F\F\F\F\F\F\F\F\F\F}{\PASS}

\stripelog{\F\F\F\F\F\F\F\F\F\F\F\F\F\F\F\F\F\F\F\F\F\F\T\F\F\F\F\F\F\F\F\F\F\F\F\F\F}{\PASS}

\stripelog{\F\F\F\F\F\F\F\F\F\F\F\F\F\F\F\F\F\F\F\F\F\F\F\F\F\T\F\F\F\F\F\F\F\F\F\F\F}{\PASS}

\stripelog{\F\F\F\F\F\F\F\F\F\F\F\F\F\F\F\F\F\F\F\F\F\F\F\F\F\F\T\F\F\F\F\F\F\F\F\F\F}{\PASS}

\stripelog{\F\F\F\F\F\F\F\F\F\F\F\F\F\F\F\F\F\F\F\F\F\F\F\F\F\F\F\T\F\F\F\F\F\F\F\F\F}{\PASS}

\stripelog{\F\F\F\F\F\F\F\F\F\F\F\F\F\F\F\F\F\F\F\F\F\F\F\F\F\F\F\F\T\F\F\F\F\F\F\F\F}{\PASS}

\stripelog{\F\F\F\F\F\F\F\F\F\F\F\F\F\F\F\F\F\F\F\F\F\F\F\F\F\F\F\F\F\T\F\F\F\F\F\F\F}{\PASS}

\stripelog{\F\F\F\F\F\F\F\F\F\F\F\F\F\F\F\F\F\F\F\F\F\F\F\F\F\F\F\F\F\F\T\F\F\F\F\F\F}{\PASS}

\stripelog{\F\F\F\F\F\F\F\F\F\F\F\F\F\F\F\F\F\F\F\F\F\F\F\F\F\F\F\F\F\F\F\F\F\F\F\T\F}{\PASS}

\stripelog{\F\F\F\F\F\F\F\F\F\F\F\F\F\F\F\F\F\F\F\F\F\F\F\F\F\F\F\F\F\F\F\F\F\F\F\F\T}{\PASS}

\stripelog{\F\F\T\F\F\F\T\T\T\T\T\T\T\T\T\T\T\T\T\T\T\T\T\F\F\T\T\T\T\T\T\F\F\F\F\T\T}{\PASS}

\stripelog{\T\T\F\F\F\F\F\T\T\T\T\T\T\T\T\T\T\T\T\T\T\T\T\F\F\T\T\T\T\T\T\F\F\F\F\T\T}{\UNRESOLVED}

\stripelog{\T\T\T\F\F\F\T\F\F\T\T\T\T\T\T\T\T\T\T\T\T\T\T\F\F\T\T\T\T\T\T\F\F\F\F\T\T}{\PASS}

\stripelog{\T\T\T\F\F\F\T\T\T\F\F\T\T\T\T\T\T\T\T\T\T\T\T\F\F\T\T\T\T\T\T\F\F\F\F\T\T}{\PASS}

\stripelog{\T\T\T\F\F\F\T\T\T\T\T\F\F\T\T\T\T\T\T\T\T\T\T\F\F\T\T\T\T\T\T\F\F\F\F\T\T}{\PASS}

\stripelog{\T\T\T\F\F\F\T\T\T\T\T\T\T\F\F\T\T\T\T\T\T\T\T\F\F\T\T\T\T\T\T\F\F\F\F\T\T}{\FAIL}

\stripelog{\T\T\T\F\F\F\T\T\T\T\T\T\T\F\F\F\T\T\T\T\T\T\T\F\F\T\T\T\T\T\T\F\F\F\F\T\T}{\PASS}

\stripelog{\T\T\T\F\F\F\T\T\T\T\T\T\T\F\F\T\F\T\T\T\T\T\T\F\F\T\T\T\T\T\T\F\F\F\F\T\T}{\FAIL}

\stripelog{\T\T\T\F\F\F\T\T\T\T\T\T\T\F\F\T\F\F\T\T\T\T\T\F\F\T\T\T\T\T\T\F\F\F\F\T\T}{\PASS}

\stripelog{\T\T\T\F\F\F\T\T\T\T\T\T\T\F\F\T\F\T\F\T\T\T\T\F\F\T\T\T\T\T\T\F\F\F\F\T\T}{\FAIL}

\stripelog{\T\T\T\F\F\F\T\T\T\T\T\T\T\F\F\T\F\T\F\F\T\T\T\F\F\T\T\T\T\T\T\F\F\F\F\T\T}{\FAIL}

\stripelog{\T\T\T\F\F\F\T\T\T\T\T\T\T\F\F\T\F\T\F\F\F\T\T\F\F\T\T\T\T\T\T\F\F\F\F\T\T}{\PASS}

\stripelog{\T\T\T\F\F\F\T\T\T\T\T\T\T\F\F\T\F\T\F\F\T\F\T\F\F\T\T\T\T\T\T\F\F\F\F\T\T}{\FAIL}

\stripelog{\T\T\T\F\F\F\T\T\T\T\T\T\T\F\F\T\F\T\F\F\T\F\F\F\F\T\T\T\T\T\T\F\F\F\F\T\T}{\PASS}

\stripelog{\T\T\T\F\F\F\T\T\T\T\T\T\T\F\F\T\F\T\F\F\T\F\T\F\F\F\T\T\T\T\T\F\F\F\F\T\T}{\PASS}

\stripelog{\T\T\T\F\F\F\T\T\T\T\T\T\T\F\F\T\F\T\F\F\T\F\T\F\F\T\F\T\T\T\T\F\F\F\F\T\T}{\FAIL}

\stripelog{\T\T\T\F\F\F\T\T\T\T\T\T\T\F\F\T\F\T\F\F\T\F\T\F\F\T\F\F\T\T\T\F\F\F\F\T\T}{\PASS}

\stripelog{\T\T\T\F\F\F\T\T\T\T\T\T\T\F\F\T\F\T\F\F\T\F\T\F\F\T\F\T\F\T\T\F\F\F\F\T\T}{\FAIL}

\stripelog{\T\T\T\F\F\F\T\T\T\T\T\T\T\F\F\T\F\T\F\F\T\F\T\F\F\T\F\T\F\F\T\F\F\F\F\T\T}{\FAIL}

\stripelog{\T\T\T\F\F\F\T\T\T\T\T\T\T\F\F\T\F\T\F\F\T\F\T\F\F\T\F\T\F\F\F\F\F\F\F\T\T}{\PASS}

\stripelog{\T\T\T\F\F\F\T\T\T\T\T\T\T\F\F\T\F\T\F\F\T\F\T\F\F\T\F\T\F\F\T\F\F\F\F\F\T}{\PASS}

\stripelog{\T\T\T\F\F\F\T\T\T\T\T\T\T\F\F\T\F\T\F\F\T\F\T\F\F\T\F\T\F\F\T\F\F\F\F\T\F}{\PASS}

\stripelog{\T\T\T\F\F\F\T\T\T\T\T\T\T\F\F\T\F\T\F\F\T\F\T\F\F\T\F\T\F\F\T\F\F\F\F\T\T}{\FAIL}

\stripelog{\T\T\T\F\F\F\T\T\T\T\T\T\T\F\F\T\F\T\F\F\T\F\T\F\F\T\F\T\F\F\T\F\F\F\F\T\T}{\PASS}

\stripelog{\F\F\T\F\F\F\T\T\T\T\T\T\T\F\F\T\F\T\F\F\T\F\T\F\F\T\F\T\F\F\T\F\F\F\F\T\T}{\PASS}

\stripelog{\T\T\F\F\F\F\F\T\T\T\T\T\T\F\F\T\F\T\F\F\T\F\T\F\F\T\F\T\F\F\T\F\F\F\F\T\T}{\PASS}

\stripelog{\T\T\T\F\F\F\T\F\F\T\T\T\T\F\F\T\F\T\F\F\T\F\T\F\F\T\F\T\F\F\T\F\F\F\F\T\T}{\PASS}

\stripelog{\T\T\T\F\F\F\T\T\T\F\F\T\T\F\F\T\F\T\F\F\T\F\T\F\F\T\F\T\F\F\T\F\F\F\F\T\T}{\PASS}

\stripelog{\T\T\T\F\F\F\T\T\T\T\T\F\F\F\F\T\F\T\F\F\T\F\T\F\F\T\F\T\F\F\T\F\F\F\F\T\T}{\PASS}

\stripelog{\T\T\T\F\F\F\T\T\T\T\T\T\T\F\F\F\F\T\F\F\T\F\T\F\F\T\F\T\F\F\T\F\F\F\F\T\T}{\PASS}

\stripelog{\T\T\T\F\F\F\T\T\T\T\T\T\T\F\F\T\F\F\F\F\T\F\T\F\F\T\F\T\F\F\T\F\F\F\F\T\T}{\PASS}

\stripelog{\T\T\T\F\F\F\T\T\T\T\T\T\T\F\F\T\F\T\F\F\F\F\T\F\F\T\F\T\F\F\T\F\F\F\F\T\T}{\PASS}

\stripelog{\T\T\T\F\F\F\T\T\T\T\T\T\T\F\F\T\F\T\F\F\T\F\F\F\F\T\F\T\F\F\T\F\F\F\F\T\T}{\PASS}

\stripelog{\T\T\T\F\F\F\T\T\T\T\T\T\T\F\F\T\F\T\F\F\T\F\T\F\F\F\F\T\F\F\T\F\F\F\F\T\T}{\PASS}

\stripelog{\T\T\T\F\F\F\T\T\T\T\T\T\T\F\F\T\F\T\F\F\T\F\T\F\F\T\F\F\F\F\T\F\F\F\F\T\T}{\PASS}

\stripelog{\T\T\T\F\F\F\T\T\T\T\T\T\T\F\F\T\F\T\F\F\T\F\T\F\F\T\F\T\F\F\F\F\F\F\F\T\T}{\PASS}

\stripelog{\T\T\T\F\F\F\T\T\T\T\T\T\T\F\F\T\F\T\F\F\T\F\T\F\F\T\F\T\F\F\T\F\F\F\F\F\T}{\PASS}

\stripelog{\T\T\T\F\F\F\T\T\T\T\T\T\T\F\F\T\F\T\F\F\T\F\T\F\F\T\F\T\F\F\T\F\F\F\F\T\F}{\PASS}

\stripelog{\T\F\F\F\F\F\F\F\F\F\F\F\F\F\F\F\F\F\F\F\F\F\F\F\F\F\F\F\F\F\F\F\F\F\F\F\F}{\PASS}

\stripelog{\F\T\F\F\F\F\F\F\F\F\F\F\F\F\F\F\F\F\F\F\F\F\F\F\F\F\F\F\F\F\F\F\F\F\F\F\F}{\PASS}

\stripelog{\F\F\T\F\F\F\F\F\F\F\F\F\F\F\F\F\F\F\F\F\F\F\F\F\F\F\F\F\F\F\F\F\F\F\F\F\F}{\PASS}

\stripelog{\F\F\F\F\F\F\T\F\F\F\F\F\F\F\F\F\F\F\F\F\F\F\F\F\F\F\F\F\F\F\F\F\F\F\F\F\F}{\PASS}

\stripelog{\F\F\F\F\F\F\F\T\F\F\F\F\F\F\F\F\F\F\F\F\F\F\F\F\F\F\F\F\F\F\F\F\F\F\F\F\F}{\PASS}

\stripelog{\F\F\F\F\F\F\F\F\T\F\F\F\F\F\F\F\F\F\F\F\F\F\F\F\F\F\F\F\F\F\F\F\F\F\F\F\F}{\PASS}

\stripelog{\F\F\F\F\F\F\F\F\F\T\F\F\F\F\F\F\F\F\F\F\F\F\F\F\F\F\F\F\F\F\F\F\F\F\F\F\F}{\PASS}

\stripelog{\F\F\F\F\F\F\F\F\F\F\T\F\F\F\F\F\F\F\F\F\F\F\F\F\F\F\F\F\F\F\F\F\F\F\F\F\F}{\PASS}

\stripelog{\F\F\F\F\F\F\F\F\F\F\F\T\F\F\F\F\F\F\F\F\F\F\F\F\F\F\F\F\F\F\F\F\F\F\F\F\F}{\PASS}

\stripelog{\F\F\F\F\F\F\F\F\F\F\F\F\T\F\F\F\F\F\F\F\F\F\F\F\F\F\F\F\F\F\F\F\F\F\F\F\F}{\PASS}

\stripelog{\F\T\T\F\F\F\T\T\T\T\T\T\T\F\F\T\F\T\F\F\T\F\T\F\F\T\F\T\F\F\T\F\F\F\F\T\T}{\PASS}

\stripelog{\T\F\T\F\F\F\T\T\T\T\T\T\T\F\F\T\F\T\F\F\T\F\T\F\F\T\F\T\F\F\T\F\F\F\F\T\T}{\FAIL}

\stripelog{\T\F\F\F\F\F\T\T\T\T\T\T\T\F\F\T\F\T\F\F\T\F\T\F\F\T\F\T\F\F\T\F\F\F\F\T\T}{\PASS}

\stripelog{\T\F\T\F\F\F\F\T\T\T\T\T\T\F\F\T\F\T\F\F\T\F\T\F\F\T\F\T\F\F\T\F\F\F\F\T\T}{\PASS}

\stripelog{\T\F\T\F\F\F\T\F\T\T\T\T\T\F\F\T\F\T\F\F\T\F\T\F\F\T\F\T\F\F\T\F\F\F\F\T\T}{\PASS}

\stripelog{\T\F\T\F\F\F\T\T\F\T\T\T\T\F\F\T\F\T\F\F\T\F\T\F\F\T\F\T\F\F\T\F\F\F\F\T\T}{\FAIL}

\stripelog{\T\F\T\F\F\F\T\T\F\F\T\T\T\F\F\T\F\T\F\F\T\F\T\F\F\T\F\T\F\F\T\F\F\F\F\T\T}{\FAIL}

\stripelog{\T\F\T\F\F\F\T\T\F\F\F\T\T\F\F\T\F\T\F\F\T\F\T\F\F\T\F\T\F\F\T\F\F\F\F\T\T}{\PASS}

\stripelog{\T\F\T\F\F\F\T\T\F\F\T\F\T\F\F\T\F\T\F\F\T\F\T\F\F\T\F\T\F\F\T\F\F\F\F\T\T}{\FAIL}

\stripelog{\T\F\T\F\F\F\T\T\F\F\T\F\F\F\F\T\F\T\F\F\T\F\T\F\F\T\F\T\F\F\T\F\F\F\F\T\T}{\PASS}

\stripelog{\T\F\T\F\F\F\T\T\F\F\T\F\T\F\F\F\F\T\F\F\T\F\T\F\F\T\F\T\F\F\T\F\F\F\F\T\T}{\PASS}

\stripelog{\T\F\T\F\F\F\T\T\F\F\T\F\T\F\F\T\F\F\F\F\T\F\T\F\F\T\F\T\F\F\T\F\F\F\F\T\T}{\PASS}

\stripelog{\T\F\T\F\F\F\T\T\F\F\T\F\T\F\F\T\F\T\F\F\F\F\T\F\F\T\F\T\F\F\T\F\F\F\F\T\T}{\PASS}

\stripelog{\T\F\T\F\F\F\T\T\F\F\T\F\T\F\F\T\F\T\F\F\T\F\F\F\F\T\F\T\F\F\T\F\F\F\F\T\T}{\PASS}

\stripelog{\T\F\T\F\F\F\T\T\F\F\T\F\T\F\F\T\F\T\F\F\T\F\T\F\F\F\F\T\F\F\T\F\F\F\F\T\T}{\PASS}

\stripelog{\T\F\T\F\F\F\T\T\F\F\T\F\T\F\F\T\F\T\F\F\T\F\T\F\F\T\F\F\F\F\T\F\F\F\F\T\T}{\PASS}

\stripelog{\T\F\T\F\F\F\T\T\F\F\T\F\T\F\F\T\F\T\F\F\T\F\T\F\F\T\F\T\F\F\F\F\F\F\F\T\T}{\PASS}

\stripelog{\T\F\T\F\F\F\T\T\F\F\T\F\T\F\F\T\F\T\F\F\T\F\T\F\F\T\F\T\F\F\T\F\F\F\F\F\T}{\PASS}

\stripelog{\T\F\T\F\F\F\T\T\F\F\T\F\T\F\F\T\F\T\F\F\T\F\T\F\F\T\F\T\F\F\T\F\F\F\F\T\F}{\PASS}

\stripelog{\T\F\T\F\F\F\T\T\F\F\T\F\T\F\F\T\F\T\F\F\T\F\T\F\F\T\F\T\F\F\T\F\F\F\F\T\T}{\PASS}

\stripelog{\F\F\T\F\F\F\T\T\F\F\T\F\T\F\F\T\F\T\F\F\T\F\T\F\F\T\F\T\F\F\T\F\F\F\F\T\T}{\PASS}

\stripelog{\T\F\F\F\F\F\T\T\F\F\T\F\T\F\F\T\F\T\F\F\T\F\T\F\F\T\F\T\F\F\T\F\F\F\F\T\T}{\PASS}

\stripelog{\T\F\T\F\F\F\F\T\F\F\T\F\T\F\F\T\F\T\F\F\T\F\T\F\F\T\F\T\F\F\T\F\F\F\F\T\T}{\PASS}

\stripelog{\T\F\T\F\F\F\T\F\F\F\T\F\T\F\F\T\F\T\F\F\T\F\T\F\F\T\F\T\F\F\T\F\F\F\F\T\T}{\PASS}

\stripelog{\T\F\T\F\F\F\T\T\F\F\F\F\T\F\F\T\F\T\F\F\T\F\T\F\F\T\F\T\F\F\T\F\F\F\F\T\T}{\PASS}


\hrulefill

\stripelog{\T\F\T\F\F\F\T\T\F\F\T\F\T\F\F\T\F\T\F\F\T\F\T\F\F\T\F\T\F\F\T\F\F\F\F\T\T}{\FAIL}

\end{multicols}
\vspace{-0.25cm}
\caption{\STRIPE run on \<sample.pl> with program output as test criterion}
\label{fig:sample-log}
\end{figure*}

\begin{figure*}
\begin{center}
\tiny
\begin{tabular}{@{}c@{\;\;}l@{\;\;}l@{\;\;}l@{\;\;}l@{}}
Event & Original trace & Reduced trace & \dots w.r.t. $\$\|sum|$ & \dots w.r.t. $\$\|mul|$ \\
\hline
$ 1_{1} $&\$\|sum| = 0;&\$\|sum| = 0;&\$\|sum| = 0;\\
$ 2_{2} $&\$\|mul| = 1;&\\
$ 3_{3} $&print \<\dq{}a? \dq{}>; $\$a = \langle\rangle$;&print \<\dq{}a? \dq{}>; $\$a = \langle\rangle$;&print \<\dq{}a? \dq{}>; $\$a = \langle\rangle$;&print \<\dq{}a? \dq{}>; $\$a = \langle\rangle$;\\
$ 4_{4} $&print \<\dq{}b? \dq{}>; $\$b = \langle\rangle$;&\\
$ 5_{5} $&\textbf{while} $(\$a \leq \$b)$ \{&\\
$ 6_{6} $&$\quad \$\|sum| = \$\|sum| + \$a$;\\
$ 7_{7} $&$\quad \$\|mul| = \$\|mul| * \$a$;&$\quad \$\|mul| = \$\|mul| * \$a$;&&$\quad \$\|mul| = \$\|mul| * \$a$\\
$ 8_{8} $&$\quad \$a = \$a + 1$;&$\quad \$a = \$a + 1$;&$\quad \$a = \$a + 1$;\\
$ 9_{9} $&\}&\\
$ 5_{10} $&\textbf{while} $(\$a \leq \$b)$ \{&\\
$ 6_{11}$&$\quad \$\|sum| = \$\|sum| + \$a$;&$\quad \$\|sum| = \$\|sum| + \$a$;&$\quad \$\|sum| = \$\|sum| + \$a$;\\
$ 7_{12}$&$\quad \$\|mul| = \$\|mul| * \$a$;&\\
$ 8_{13}$&$\quad \$a = \$a + 1$;&$\quad \$a = \$a + 1$;&$\quad \$a = \$a + 1$;\\
$ 9_{14}$&\}&\\
$ 5_{15}$&\textbf{while} $(\$a \leq \$b)$ \{&\\
$ 6_{16}$&$\quad \$\|sum| = \$\|sum| + \$a$;&$\quad \$\|sum| = \$\|sum| + \$a$;&$\quad \$\|sum| = \$\|sum| + \$a$;\\
$ 7_{17}$&$\quad \$\|mul| = \$\|mul| * \$a$;&\\
$ 8_{18}$&$\quad \$a = \$a + 1$;&$\quad \$a = \$a + 1$;&$\quad \$a = \$a + 1$;\\
$ 9_{19}$&\}&\\
$ 5_{20}$&\textbf{while} $(\$a \leq \$b)$ \{&\\
$ 6_{21}$&$\quad \$\|sum| = \$\|sum| + \$a$;&$\quad \$\|sum| = \$\|sum| + \$a$;&$\quad \$\|sum| = \$\|sum| + \$a$;\\
$ 7_{22}$&$\quad \$\|mul| = \$\|mul| * \$a$;&\\
$ 8_{23}$&$\quad \$a = \$a + 1$;&$\quad \$a = \$a + 1$;&$\quad \$a = \$a + 1$;\\
$ 9_{24}$&\}&\\
$ 5_{25}$&\textbf{while} $(\$a \leq \$b)$ \{&\\
$ 6_{26}$&$\quad \$\|sum| = \$\|sum| + \$a$;&$\quad \$\|sum| = \$\|sum| + \$a$;&$\quad \$\|sum| = \$\|sum| + \$a$;\\
$ 7_{27}$&$\quad \$\|mul| = \$\|mul| * \$a$;&\\
$ 8_{28}$&$\quad \$a = \$a + 1$;&$\quad \$a = \$a + 1$;&$\quad \$a = \$a + 1$;\\
$ 9_{29}$&\}&\\
$ 5_{30}$&\textbf{while} $(\$a \leq \$b)$ \{&\\
$ 6_{31}$&$\quad \$\|sum| = \$\|sum| + \$a$;&$\quad \$\|sum| = \$\|sum| + \$a$;&$\quad \$\|sum| = \$\|sum| + \$a$;\\
$ 7_{32}$&$\quad \$\|mul| = \$\|mul| * \$a$;&\\
$ 8_{33}$&$\quad \$a = \$a + 1$;&\\
$ 9_{34}$&\}&\\
$ 5_{35}$&\textbf{while} $(\$a \leq \$b)$ \{&\\
$10_{36}$&print \<\dq{}sum = \dq{}>, \$\|sum|, \<\dq{}$\backslash$n\dq{}>;&print \<\dq{}sum = \dq{}>, \$\|sum|, \<\dq{}$\backslash$n\dq{}>;&print \<\dq{}sum = \dq{}>, \$\|sum|, \<\dq{}$\backslash$n\dq{}>;\\
$11_{37}$&print \<\dq{}mul = \dq{}>, \$\|mul|, \<\dq{}$\backslash$n\dq{}>;&print \<\dq{}mul = \dq{}>, \$\|mul|, \<\dq{}$\backslash$n\dq{}>;&&print \<\dq{}mul = \dq{}>, \$\|mul|, \<\dq{}$\backslash$n\dq{}>;
\end{tabular}
\end{center}
\caption{Execution traces of \<sample.pl>}
\label{fig:sample-traces}
\end{figure*}

\pagebreak
\noindent
In our example, the $\test$ function would return
\begin{itemize}
\item $\FAIL$ whenever the expected behavior was actually reproduced,
\item $\PASS$ when anything else was produced, and
\item $\UNRESOLVED$ if the program did not produce anything, i.e. the
  program crashed or hung.
\end{itemize}

\noindent
Figure~\vref{fig:sample-log} shows the \STRIPE run.  \STRIPE required
a total of 176~tests (i.e.\ partial executions) to reduce the
execution trace.  Each bullet $\bullet$ stands for an
applied~$\Delta_i$, or executed statement.  We see that most tests
either result in $\PASS$~or~$\UNRESOLVED$.  Only after 50~tests do we
see a $\FAIL$~test outcome---the expected output was produced and the
trace can be reduced to the shown set of statements.  The last line
shows the final 1-minimal trace.  The actual result is shown in
Figure~\ref{fig:sample-traces}.  All events are listed in the form
$\|line number|_{\|time|}$.

\begin{itemize}
\item The original trace is shown in the first column; we see how the
  variables $\$a$~and~$\$b$ are read in (events $3_3$~and~$4_4$ in
  Figure~\ref{fig:sample-log}) and how $\$\|sum|$~and~$\$\|mul|$ are
  computed and printed.
  
\item The reduced trace in the second column shows that several of
  these statements are actually irrelevant for computing the
  output---executing the original trace and the reduced trace has the
  same effect.  Thus, we find that the initialization of
  $\$\|mul|$ ($1_1$) is irrelevant, since whatever it is initialized
  to, $\$\|mul|$ will always be zero.\footnote{Likewise, all later
    assignments to $\$\|mul|$ ($7_{12}, 7_{17}, 7_{22}, \dots$) do not
    change its value.  Finally, all control statements are irrelevant,
    since the control flow is explicitly stated in the execution
    trace.}

\item In the third column, we have run \STRIPE with a $\test$ function
  that compares only the $\$\|sum|$ output.  This means that the few
  statements related to $\$\|mul|$ can be eliminated as well.
  
\item In the fourth and last column, we have run \STRIPE with a
  $\test$ function that compares only the $\$\|mul|$ output.  Only the
  initial assignment and the final output remain.
\end{itemize}

\noindent
We see how \STRIPE effectively reduces execution traces to those
events which were actually relevant (or \emph{critical}) in reaching
the final state.  We thus call this sequence of events a
\emph{critical slice}---similar to the critical slices as explored by
DeMillo, Pan and Spafford~\cite{demillo/pan/spafford/96/issta}, but
guaranteeing 1-minimality and with a much better best-case efficiency.
Furthermore, delta debugging does not require any program analysis (in
the case of \PERL programs such as \<sample.pl>, this is an almost
impossible issue, anyway).

\section{Conclusion and Future Work}

Delta debugging automates the most time-consuming debugging issue:
determining the \emph{relevant problem circumstances.}  Relevant
circumstances include the program input, changes to the program code,
or executed statements.  All that is required is an automated
test.

Delta debugging comes at a price: Although the $\ddmin$ algorithm
guarantees 1-minimality, the worst-case quadratic complexity is a
severe penalty for real-world programs---especially considering
program runs with billions of executed statements.  Consequently, our
future work will concentrate on introducing \emph{domain knowledge}
into delta debugging.  In the domain of code changes, we have seen
significant improvements by grouping changes according to files,
functions, or static program slices, and rejecting infeasible
configurations; we expect similar improvements for program input and
program statements.

Our long-term vision is that, to debug a program, all one has to do is
to set up an appropriate $\test$ function.  Then, one can let the
computer do the debugging, isolating failure circumstances using a
combination of program analysis and automated testing.  Automatic
isolation of failure causes is no longer beyond the state of the
art.  It is just a question of how much computing power and 
program analysis you are willing to spend on it.
\medskip

\noindent
\textbf{Acknowledgements.}  Ralf Hildebrandt, Kerstin Reese, and
Gregor Snelting provided valuable comments on earlier revisions of
this paper.
\medskip

\noindent
Further information on delta debugging is available at
\begin{center}
\texttt{http://www.fmi.uni-passau.de/st/dd/}\enspace.
\end{center}

\end{document}